\documentclass{article}

\usepackage{PRIMEarxiv}

\usepackage[utf8]{inputenc} 
\usepackage[T1]{fontenc}    
\usepackage{hyperref}       
\usepackage{url}            
\usepackage{booktabs}       
\usepackage{amsfonts}       
\usepackage{nicefrac}       
\usepackage{microtype}      
\usepackage{amsmath}        
\usepackage{lipsum}
\usepackage{fancyhdr}       
\usepackage{graphicx}       
\graphicspath{{media/}}     

\title{Integrated effect of the cosmic space magnetic field on the acceleration noise of the TQ gravitational wave detection program
\thanks{\textit{\underline{Citation}}: 
\textbf{Authors. Title. Pages.... DOI:000000/11111.}} 
}

\author{
  Cheng-Long Yu \\
  Jin Yan \\
  Lin Ji \\
  Wen-Ke Shi \\
  Yi Zhang \\
  Run-Qiu Liu \\
Hong-Qing Huo\thanks{*Corresponding Author. E-mail: huohq@lzu.edu.cn  ;  Lanzhou University, Lanzhou , China.}  
}

\begin{document}
\maketitle

\begin{abstract}

The TianQin(TQ) program is to deploy three satellites that can form an equilateral triangle in about 100,000 km Earth orbit to capture gravitational wave signals in the low-frequency band. In order to ensure accurate capture, noise needs to be analyzed and compensated. In this paper, we model and analyze the acceleration noise generated by the test mass affected by the magnetic field in space. In this paper, we use the Tsyganenko model as the background magnetic field of the TQ orbit, calculate the magnetic field and magnetic field gradient of the satellite orbit from 1997 to 2023, analyze the acceleration noise due to the coupling of the residual magnetic moment, the induced magnetic moment with the magnetic field in space and the acceleration noise due to the Lorentz force, and calculate the acceleration integrated noise of the influence of the magnetic field on the test mass from the power spectral densities of the modeled magnetic field and the magnetic field gradient. The acceleration integrated noise of the magnetic field influence on the test mass is calculated from the power spectral density of the magnetic field and the magnetic field gradient obtained by the model. Through the simulation study, the acceleration of the test mass induced by the magnetic field in the space of the TQ orbit reaches the magnitude of $10^{-16}ms^{-1}Hz^{-1/2}$, which is an important source of the influencing noise. The acceleration noise induced by the magnetic field and the Lorentz force is relatively higher than that induced by the magnetic field gradient.
\end{abstract}

\keywords{cosmic space magnetic field \and acceleration noise \and gravitational wave detection}

\section{Introduction}

 At present, there are many types of gravitational wave detectors\cite{luo2019overall}, including ground-based laser interferometers, space laser interferometers, pulsar timing arrays and so on. The ground-based gravitational wave detection program mainly detects gravitational wave signals in the frequency band of 10Hz-1kHz, the space-based gravitational wave detection program mainly detects gravitational wave signals in the frequency band of 0.1mHz-1Hz, and the Pulsar Timing Array detects even lower frequency bands in the nanohertz band. Ground-based gravitational wave detection programs include LIGO (Laser Interferometric Gravitational Wave Observatory)\cite{Aasi_2015}, Virgo (Virgo Gravitational Wave Observatory)\cite{PhysRevX.9.031040}, KAGRA (Kamioka Gravitational Wave Detector)\cite{10.1093/ptep/ptaa125}, etc.; space-based gravitational wave detection programs include LISA led by ESA\cite{auclair2023cosmology}, China's TQ Program\cite{torres2024detection} and Taiji Program\cite{doi:10.1142/S0217751X2050075X}, and Japan's DECIGO\cite{10.1093/ptep/ptab019}. So far, more than a hundred cases of gravitational wave signals have been detected, most of which are high-frequency (10-1000Hz) signals detected by ground-based gravitational wave detectors such as LIGO, Virgo, and KAGRA, while the primary goal of the space-based gravitational wave detection program is to detect gravitational wave signals in the low-frequency (0.1mHz-1Hz) range.
 
 The space gravitational wave detector is a giant interferometer\cite{s24237685}, where two test masses, far apart, undergo a slight change in displacement when a gravitational wave passes through, and in order for the detector to detect this change, the acceleration produced by other noisy forces must be brought up to a very high level. As energetic particles from galactic and solar system cosmic rays penetrate the outer wall of the spacecraft\cite{https://doi.org/10.1029/2020JA028579}, the test mass becomes charged, and these charges interact with the magnetic field in space, generating the Lorentz force, which produces the acceleration noise that interferes with the detection. At this point, the capacitive sensors near the test mass measure the relative change in displacement and pass this information to the Newtonian thrusters, compensating for the noise of this item, so that the test mass does free-fall motion free from external interference. In order to achieve this goal, we must be clear about the magnetic field of the orbit in which the satellite is located and deduce the effect of the magnetic field on the test mass. In this paper, the magnetic field environment in space is investigated using the Tsyganenko model\cite{https://doi.org/10.1029/93JA01150,TSYGANENKO19895}, and numerical simulation and analysis of the magnetic field in the orbit of the TQ program are focused.
 
 The TQ program, a space-based gravitational wave detection program proposed in 2014\cite{gong2021concepts}, consists of three satellites deployed in an orbit of 100,000 km to form an equilateral triangle, with the satellites 170,000 km apart, using laser interferometry to capture pico-scale distance variations caused by gravitational waves\cite{Su_2021}. These gravitational waves may come from astrophysical events such as supermassive black hole mergers and extreme mass ratio spin-ups. Therefore, the requirements of the TQ program for acceleration noise are of the order of$1\times 10^{-15}\:m\cdot s^{-2}\cdot Hz^{-1/2}$ in the sensitive frequency band\cite{Luo_2016,10.1093/ptep/ptaa114}, and the magnetic field environment-related coupling noise is an important noise perturbation source.
 
 In Su's work\cite{Su_2020}, they established a global magnetohydrodynamic (MHD) simulation in 2020 to simulate the Earth's magnetospheric structure using the SWMF model, which computed the space magnetic field environment of the TQ orbit. They calculated the acceleration noise of the space magnetic field environment for the test mass. However, due to insufficient model resolution (the minimum grid size of the SWMF model is 0.25 Earth radii), which can lead to underestimation of high-frequency noise, in 2023, they improved the model\cite{PhysRevD.108.103030}. The space magnetic field used the empirical model, the Tsyganenko model, to simulate the magnetic field environment on the TQ orbits from 1998 to 2020, and analyzed the amplitude spectrum of the acceleration noise (ASD) distribution. At the same time, they also added a risk assessment index to assess the probability of acceleration noise exceeding the standard by taking the ratio of the calculated acceleration noise to the acceleration noise required by TQ as the index R. They calculated that at the 1 mHz frequency point, the mean value of R is at 0.123, with a probability of 7.9\% for greater than 0.23, and a probability of 1.2\% for greater than 0.3; at the 6 mHz frequency point, the mean value of R is at 0.027, never exceeding 0.2. In their calculations, only the effect of the magnetic field on the total test mass is given, and the acceleration noise spectrum from$1\times 10^{-5}$ Hz to $1\times 10^{-2} $Hz is given, missing the 0.01 Hz to 0.1 Hz band noise.
 
 In this paper, using Tsyganenko's improved TA16 model from 2016\cite{https://doi.org/10.1002/2016JA023217}, the effect of the magnetic field on the test quality is separated into components of the magnetic field, the magnetic field gradient, and the Lorentz force, and the magnitude of the acceleration noise they produce is compared. In addition, the frequency domain range of the acceleration noise spectrum is increased to the range of $1\times 10^{-5}$Hz to $1\times 10^{-1}$Hz, which makes the noise spectrum in the lower frequency band more informative. In this paper, we use continuous long series data to calculate the global ASD from 1997 to 2021, which will directly respond to the spectral characteristics of the whole period, such as the correlation between orbits and the effect of the solar cycle, etc., and retain all the low-frequency components to avoid information loss. The simulation results show that the highest noise is lower than $1\times 10^{-14}\:m\cdot s^{-2}\cdot Hz^{-1/2}$ in the low-frequency band and lower than $1\times 10^{-16}\:m\cdot s^{-2}\cdot Hz^{-1/2}$ in the middle and high frequency bands.

\section{Theoretical analysis}

\subsection{Tsyganenko model}

The TQ program's satellite orbit\cite{milyukov2020tianqin} is the Earth's central orbit. Three all-synchronous satellites are deployed in an orbit of 100,000 km to form an equilateral triangle constellation with a side length of about 170,000 km. The constellation's orbital plane is roughly perpendicular to the ecliptic plane, with an orbital period of about 3.9 days. Most of this orbit lies in the Earth's magnetosphere, so the model chosen for this paper should be based on the Earth's magnetospheric magnetic field model.

Regarding the Earth's magnetospheric magnetic field models, they are divided into many types, there are empirical models, such as the Tsyganenko series models and the Mead-Fairfield model; physical models, which are mainly based on the principles of magnetohydrodynamics (MHD) and other principles, and simulate the changes of plasma and magnetic field in the magnetosphere by numerically solving the control equations; and hybrid models, which are generally empirical models plus the magnetospheric boundary conditions, and then a physical model to simulate the magnetic field variations within the magnetosphere. The empirical model is data-driven, can accurately reproduce the historical magnetic field structure, and uses parametric formulas for fast computation, with the disadvantage of poor extrapolation in the undetected region of the satellite. Physical models describe the magnetic field with precise physical mechanisms and have good extrapolation capability, but are computationally inefficient and time-consuming due to the complexity of the equations. Since most of our selected orbital regions are detected by satellites, and the time range we need to calculate is extensive, we choose the Tsyganenko model in this paper. The Tsyganenko model is a series of empirical models for describing the Earth's magnetospheric magnetic field, whose core idea is to use the theoretical formulas of the satellite detection data to construct a parameterized model of the dynamic Earth's magnetospheric magnetic field. The models focus on the effects of the interstellar magnetic field (IMF), solar wind-related parameters, and geomagnetic perturbation indices on the magnetosphere. The Tsyganenko models are empirical models based on satellite observations and physical models. After many years of updating and iterating, the Tsyganenko models can describe complex geomagnetic activities, such as the magnetotailed plasma sheet and the magnetic field description of the near-Earth radiation belt. In modelling the Tsyganenko model, not only are the magnetic field variations under different solar wind conditions taken into account, but also the tilting of the geomagnetic axis and the influence of the penetration field in the polar region.

 The model used in this paper is the TA16 model. The central idea is to divide the magnetic field into two parts, the toroidal and poleward fields, and then expand each part into a weighted sum of radial basis functions (RBF). The toroidal field is generated by the toroidal potential function$\Psi_{1}$ , which primarily describes the magnetic field components that are closed around the Earth, such as the ring currents, and the polar field is generated by the polar potential function $\Psi_{2}$ , which describes the magnetic field components that run from the poles to the equator, such as the magnetopause currents. The expression for the magnetic field synthesized by these two terms is\cite{https://doi.org/10.1029/RG014i002p00199}
 
\begin{equation}
 B=\nabla\times(\Psi_{1} r)+\nabla\times\nabla\times(\Psi_{2} r)
 \label{eq1}  
\end{equation}
    
 The toroidal potential function and the poleward potential function can mainly be expanded to form a linear combination of RBFs
 
\begin{equation}
\Psi_{1,2}(r)=\sum_{i=1}^{N} a_{i_{1,2}}\chi_{i}(r)
 \label{eq2}  
\end{equation}	
        
 Regarding the grid design of the RBF, TA16 adopts a nine concentric spheres with a distance range of 3.3Re-14.5Re, which is just enough to include the near-earth data and the magnetotail data. At the same time, TA16 adopts different node densities; the node spacing in the near-earth region, due to the drastic changes in the magnetic field, is about 0.7Re. In contrast, in the far-magnetotail region where the magnetic field gradient is flat, the spacing adopted is about 3Re, which can effectively describe the structure of the magnetosphere and adapt to the magnetic field gradient changes.
 
 Regarding the dipole tilt effect of geomagnetism\cite{https://doi.org/10.1029/2012JA018056}, the dipole tilt leads to the bending of the current sheet, and dynamic nodes adjust the model. Also, the nodes must be symmetrically distributed to reflect the symmetry of the Earth's magnetic field\cite{TSYGANENKO19871347}. The generating function of the magnetic field can be combined with the trigonometric function of the dipole tilt angle. In TA16, a term dealing with IMF By penetration effects is also added. The toroidal field part is combined with odd-universal symmetry to generate the transverse magnetic field component, which is positively correlated with the IMF $B_{y}$, and the polar field part is combined with even-universal symmetry to affect the standard component, which responds to the modulation of the current sheet morphology by the $B_{y}$.
 
 The TA16 model solves the systematic bias of traditional models (e.g., T02\cite{https://doi.org/10.1029/2001JA000219,https://doi.org/10.1029/2001JA000220}, TA15\cite{https://doi.org/10.1002/2015JA021641}) in the $B_{y}$ component. The model successfully reproduces the synchronous orbital magnetic field variations, such as $B_{z}$ pulses induced by a sudden increase in solar wind pressure\cite{BORODKOVA20081220}. For the first time, the global magnetospheric model is constructed based on a purely data-driven (no preset current module) approach, which verifies the feasibility of the RBF method in complex magnetic field modelling.

\subsection{Acceleration noise modelling}

 It is well known that the universe is full of plasma, the solar wind carries high-energy particles to bombard the spacecraft's surface, and a part of the high-energy particles passes through the shell of the spacecraft and enters its interior. The particles hit the test mass, which makes the test mass electrically charged. The charged test mass interacts with the magnetic field in space, generating acceleration noise. The residual magnetic moment of the test mass also couples with the magnetic field and interacts with the magnetic field gradient, generating additional acceleration noise. These two parts of the noise are calculated separately.

 \subsubsection{Coupling of spatial magnetic field and magnetic moment}
 
  As the magnetic moment of the test mass will not be entirely equal to zero after processing and manufacturing, a specific residual magnetic moment exists, which interacts with the spatial magnetic field and generates a magnetic force. Under the influence of the spatial magnetic field, the test mass also generates an induced magnetic moment, which interacts with the spatial magnetic field and generates acceleration noise.
  
 In this paper since the scale of the size of the test mass is much smaller than the scale of the displacement change of the spacecraft at the minimum rate of change in time, we consider the test mass as a dipole with mass m with a residual magnetic moment $M_{0}$ , and the external magnetic field is denoted by B . The following equation can express the force on the dipole in the external magnetic field B:
 
\begin{equation}
F=(M_{0}\cdot\nabla)B
 \label{eq3}  
\end{equation}

 The magnetic moment induced by the test mass in the external magnetic field $M_{ind}$ is determined by the magnetization of the material $\chi$, the volume and the external magnetic field B.
 
\begin{equation}
M_{ind}=\frac{\chi VB}{\mu_{0}}
 \label{eq4}  
\end{equation}
        
 The induced magnetic moment also interacts with the external magnetic field and generates additional forces.
 
\begin{equation}
F_{ind}=(\frac{\chi VB}{\mu_{0}}\cdot\nabla)B
 \label{eq5}  
\end{equation}
    
 Therefore, the total force received by the test mass is the sum of the forces generated by the residual and induced magnetic moments.
 
\begin{equation}
F_{total}=(M_{0}\cdot\nabla)B+\frac{\chi V}{\mu_{0}}[(B\cdot\nabla)B]
 \label{eq6}  
\end{equation}
    
 Since the magnetic field B and its gradient may not be uniform within the TM volume, volume averaging is required, specifically
 
\begin{equation}
F_{total}=<(M_{0}\cdot\nabla)B+\frac{\chi V}{\mu_{0}}[(B\cdot\nabla)B]>
 \label{eq7}  
\end{equation}
    
 where <...> is defined as $\iiint ... \, d^{3}x$ \cite{PhysRevD.97.122002}. Written in the form of acceleration as
 
\begin{equation}
a=\frac{1}{m}<(M_{0}\cdot\nabla)B+\frac{\chi V}{\mu_{0}}[(B\cdot\nabla)B]>
 \label{eq8}  
\end{equation}
        
 Since there are electrode cages and inertial sensor housings outside the test mass, these metal structures will provide some magnetic shielding effect,$\eta$, and here we take a shielding leakage coefficient, , and add it to the acceleration equation, representing the magnetic shielding effect of all the devices outside the test mass.
 
\begin{equation}
a=\eta\frac{1}{m}<(M_{0}\cdot\nabla)B+\frac{\chi V}{\mu_{0}}[(B\cdot\nabla)B]>
 \label{eq9}  
\end{equation}
        
 Under reasonable assumptions —— homogeneity and stability,we can convert the acceleration formula above into a power spectral density. Here comes, we are only concerned with the noise in the direction of the sensitive axis, the
 
\begin{equation}
S_{a,x}=(\frac{\eta\chi V}{m\mu_{0}}<\nabla B_{x}>)^{2}S_{B_{x}}+(\frac{\eta}{m}<M_{0}>+\frac{\eta\chi V}{m\mu_{0}}<B_{x}>)^{2}S_{\nabla B_{x}}
 \label{eq10}  
\end{equation}
        
 S stands for the power spectral density. The first term is the contribution of the magnetic field to the acceleration noise, and the second term is the effect of the magnetic field gradient on the acceleration noise. In LISApathfinder it is considered that the first term of the above equation is dominant in the low-frequency region because of the coupling of the magnetization of the test mass,$\chi$ , to the interstellar magnetic field, $\nabla B_{x}$; the second term is the noise induced by the gradient of the magnetic field, which has minimal effect in the low-frequency band but shows a dominant tendency in the high-frequency band. The results of our calculations for the above two terms will be shown in the next section.

\subsubsection{Lorentz force noise equation}

When solar particles and high-energy rays reach the satellite orbit, they penetrate the test mass shell and hit its surface, electrifying the test mass. Under the influence of the background magnetic field, the charged particles are subjected to the Lorentz force in space, resulting in additional acceleration noise.The metal shell of the test mass will produce a shielding effect under the influence of the Hall voltage in the space magnetic field, with the same shielding leakage coefficient $\eta$ as mentioned above. The acceleration of the test mass due to the Lorentz force in space is\cite{Sumner_2020}

\begin{equation}
a_{L}=\frac{Q}{m}(\eta v_{SC}B+v_{PM}(B+B_{int}))
 \label{eq11}  
\end{equation}
        
 Here Q is the charge of the test mass, m is the mass of the test mass, $\eta$is the shielding coefficient of the external magnetic field,$v_{SC}$ is the speed of the spacecraft concerning the external magnetic field, $v_{PM}$ is the speed of the test mass concerning the spacecraft, B is the external magnetic field of the spacecraft, and $B_{int}$ is the internal magnetic field generated by various instruments inside the spacecraft. All the variables related to $a_{L}$ in this equation,Q,$v_{SC}$,$v_{PM}$,B,and $B_{int}$ may generate noise, but here we are only concerned with the effect of the external magnetic field on the test mass, so we only analyze B here. The acceleration noise generated by B, can be expressed by the following equation.
 
\begin{equation}
S_{a_{L}}=[\frac{Q}{m}(\eta v_{SC}+v_{PM})]^{2}S_{B}
 \label{eq12}  
\end{equation}
     
 The above equation is the power spectral density. In general, we use the amplitude spectral density.
 
\begin{equation}
S^{1/2}_{a_{L}}=|\frac{Q}{m}(\eta v_{SC}+v_{PM})|S^{1/2}_{B}
 \label{eq13}  
\end{equation}
        
 This gives us the spectral density of the acceleration noise. In the next section, we will calculate the spectral density of the external magnetic field, after which we can get the noise spectral density of the acceleration by incorporating the above equation.
 
\subsection{Orbital magnetic field modelling}

 The Tsyganenko model uses magnetic field data from Geotail, Polar, and Cluster satellites, and also combines THEMIS/ARTEMIS, Van Allen, and other related data, modelling the region from 2-60 Re, and further restricting the magnetotail region, $x_{gsw}$, to -20 Re. The validation set uses data from the GOES 10 satellite. The driving parameters include the solar wind pressure Pdyn, the y-component of the interstellar magnetic field IMF $B_{y}$, and the geomagnetic perturbation index, including the corrected Sym-H parameter with sliding mean and time derivative, Sym-Hc, the N-index that quantifies the efficiency of the solar wind-magnetosphere coupling, and other parameters.
 
 In this paper, we take an orbit perpendicular to the ecliptic plane, with an orbital altitude of 100,000 km and a satellite operational period of 3.9 days for a circular orbit, and we take the data from 1997 to 2021 as the period for our analysis. Every eleven years represents one solar cycle, so here we take about two solar cycles of data.
 
 Regarding the TA16 model, its time resolution is 5min, while the spectrum we need to calculate ranges from $1\times 10^{-5}$Hz to 0.1Hz, and the time resolution of 5min is far from the maximum frequency we need. Hence, we need to perform the interpolation operation here. The specific operation uses the TA16 model to calculate the magnetic field of the time domain data, using spline interpolation. This paper chooses to interpolate to a time resolution of 1s to ensure that it reaches the frequency of 0.1Hz and can better know the spectral changes in the vicinity of 0.1Hz.
 
 To find the coupling between the residual magnetic moment and the magnetic field, we also need to use the TA16 model to find the magnetic field gradient. The gradient formula combines the fourth-order centre difference formula and Richardson extrapolation.
 
 The fourth-order central difference formula uses the weighted average of the function values of five neighbouring points to approximate the derivative with an error of $O(h^{4})$, and the specific formula takes the form of
 
\begin{equation}
f'(x_{0})\approx \frac{-f(x+2h)+8f(x+h)-8f(x-h)+f(x-2h)}{12h}
 \label{eq14}  
\end{equation}
        
 The formula eliminates Taylor expansion's lower-order error terms and retains only the $h^{4}$ order error.
 
 The core idea of Richardson's extrapolation method is to eliminate the lower-order error terms by linearly combining the results of calculations with different step sizes. Using the above fourth-order central difference formula, the results of $D_{0}(h)$ and $D_{0}(\frac{h}{2})$ with step sizes h and $\frac{h}{2}$ are used to construct  Eq.\ref{eq14}.

\begin{equation}
D_{1}(h)=\frac{16D_{0}(\frac{h}{2})-D_{0}(h)}{15}
 \label{eq15}  
\end{equation}
        
 This operation eliminates the $h^{4}$ order error and improves the accuracy to $O(h^{6})$.

\section{Results and Discussion}

\subsection{Results}

\begin{figure}
  \centering
  \includegraphics[width=0.6\textwidth]{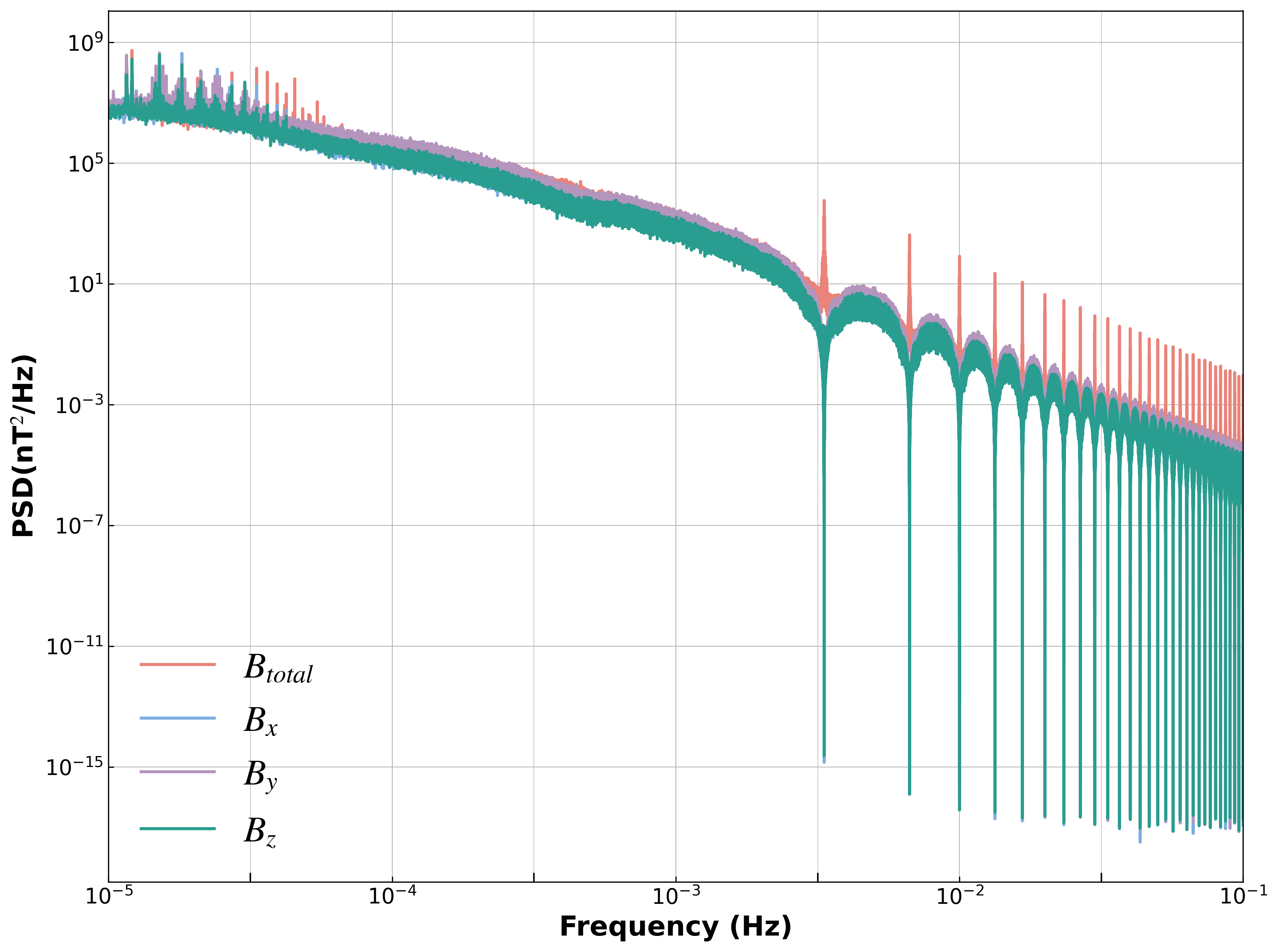}
  \caption{Power spectral density of the three components of the magnetic field calculated by TA16 and the synthesized total magnetic field}
  \label{fig:fig1}
\end{figure}

We first calculate the power spectral density of the magnetic field with all the parameters in Table \ref{tab:table}. The power spectral density of the magnetic field in the GSE coordinate system is shown in Fig. \ref{fig:fig1}, where we calculate the power spectral densities of the three components of the magnetic field, $B_{x}$, $B_{y}$, and $B_{z}$, and of the total magnetic field, B, respectively, in the TQ orbit. We can see that the PSD of the magnetic field is larger at low frequencies. The higher the frequency is, the PSD of the magnetic field decreases gradually, which is characterized by a power-law distribution. A pronounced energy concentration area in the low-frequency band reflects the geomagnetic activity and space weather events. At about $5\times 10^{-2}$Hz, the value of PSD decreases to $1 nT^{2}\cdot Hz^{-1}$, and we find that the calculated results agree with the magnetospheric turbulence spectrum. The PSD in the high-frequency part shows a tendency to oscillate, which is due to the high-frequency oscillation phenomenon because of the interpolation process to improve the frequency range of the PSD during the calculation of the magnetic field.

\begin{table}
 \caption{Parameters and Values\cite{PhysRevD.97.122002}}
  \centering
  \begin{tabular}{lll}
    \toprule
    \cmidrule(r){1-2}
    Parameter     & Values  \\
    \midrule
    $\chi$ & $-3.3723\times 10^{-5}$\cite{PhysRevD.97.122002}   \\
    $M_{x}$     &  $0.140\:nAm^{2}$ \cite{PhysRevD.97.122002}    \\
    $M_{y}$     &  $0.178\:nAm^{2}$ \cite{PhysRevD.97.122002} \\
    $M_{z}$     &  $0.095\:nAm^{2}$ \cite{PhysRevD.97.122002}   \\
    $Q$     &  $1\times 10^{-12}C$  \\
    $M$     &  $1.928\:kg$  \cite{Sumner_2020}\\
    $v_{PM}$     &  $10\pi f^{3/2} \:m/s$  \cite{Sumner_2020}\\
    $v_{SC}$     &  $3\times 10^{4}\:m/s$  \cite{Sumner_2020}\\
    \bottomrule
  \end{tabular}
  \label{tab:table}
\end{table}

\begin{figure}
  \centering
  \includegraphics[width=0.6\textwidth]{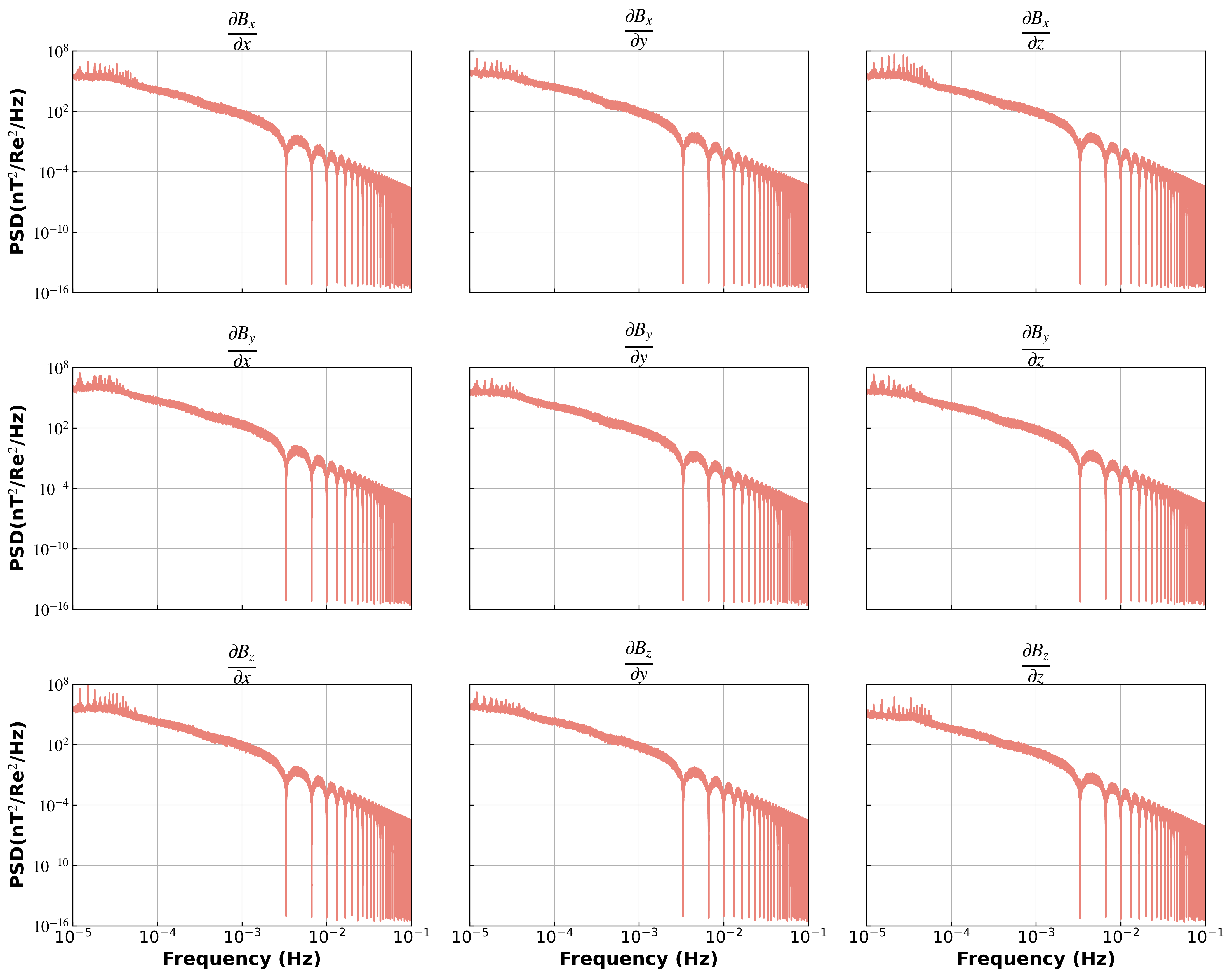}
  \caption{Power spectral density of the magnetic field gradient tensor in the orbit where the TQ program is located, with nine plots representing the PSDs of each of the nine components.}
  \label{fig:fig2}
\end{figure}

 After calculating the magnetic field of the orbit using the TA16 model, we take h = 0.3Re, and using Eq.\ref{eq14} and Eq.\ref{eq15}, we can calculate the gradient tensor of the magnetic field of the orbit where the celestial organ is located. We calculate the PSD for each of the nine components individually, as shown in Fig.\ref{fig:fig2}. The PSDs of these nine components are all the highest in the low-frequency region $1\times 10^{-5}Hz$, and we find that the decreasing tendency of the magnetic field gradient with frequency is more significant than that of the magnetic field. The values of each component are not the same, which reflects the anisotropic nature of the gradient field, and further reflects the complex structure of the magnetosphere, such as the directionality of the current sheet at the top of the magnetosphere structure of the current sheet at the top of the magnetosphere.

\begin{figure}
  \centering
  \includegraphics[width=0.6\textwidth]{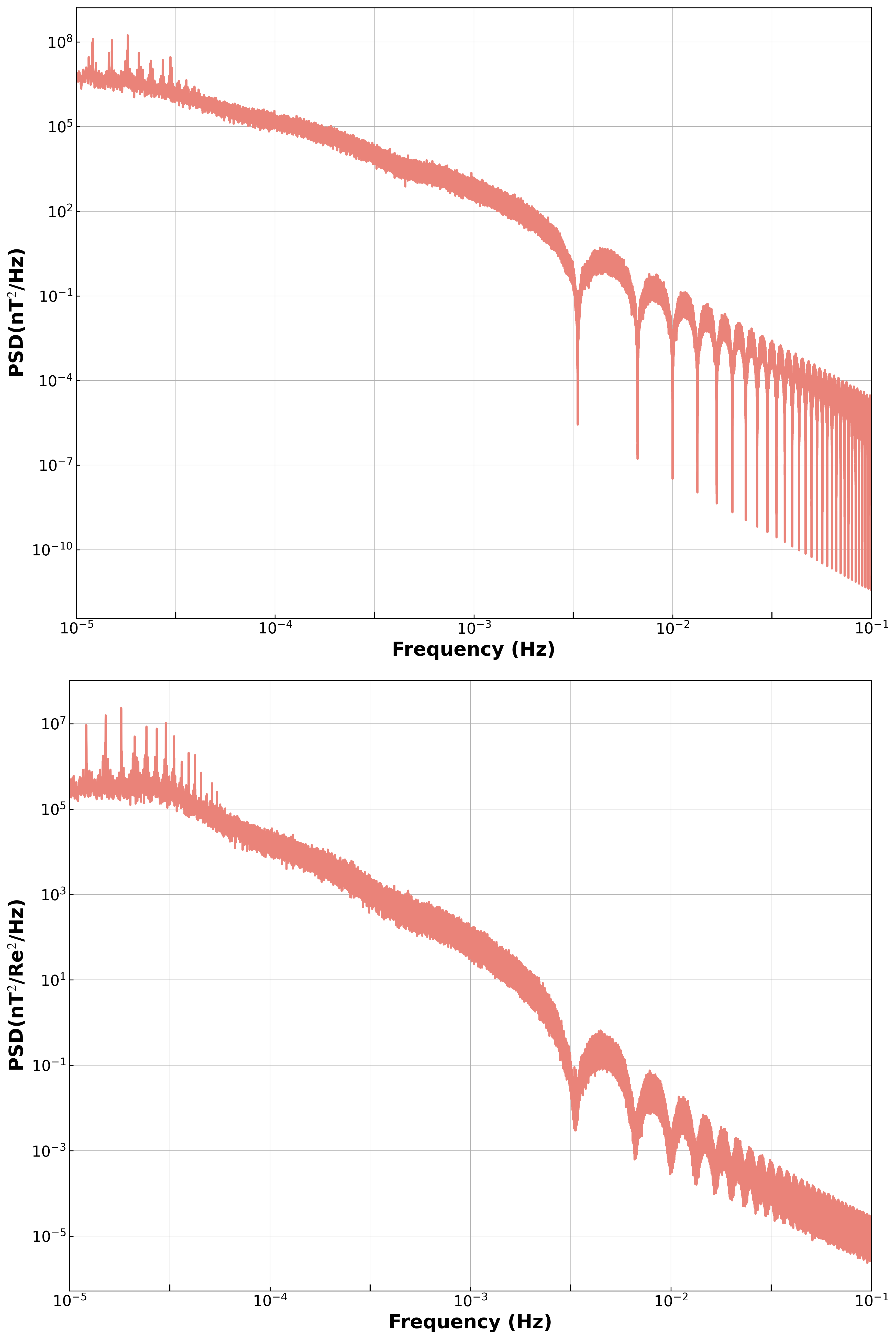}
  \caption{Power spectral densities of the magnetic field and magnetic field gradient in the sensitive axis on the celestial TQ orbit
 (The upper figure shows the power spectral density of the magnetic field, and the lower figure shows the power spectral density of the magnetic field gradient)}
  \label{fig:fig3}
\end{figure}

 For the space gravitational wave detection program, the detection of the sensitive axis direction has the most stringent requirements for the noise perturbation demand, so we need to analyze and calculate the effect of the magnetic field and gradient field in the sensitive axis direction based on the above components of the magnetic field and gradient field, as shown in Fig.\ref{fig:fig3}. As can be seen from the figure above, the PSD of the magnetic field in the sensitive axis has almost the same trend as the PSD of the magnetic field in the three components as calculated in Fig.\ref{fig:fig1} above, but in the low-frequency portion of we will find a slightly fluctuating PSD situation, because the sensitive axis is changing with time, which is equivalent to adding an oscillating source in the low-frequency part, all the fluctuations can be seen in the low-frequency part, and it is also evident from the right figure that there is an oscillating phenomenon of PSD in the low-frequency part, which is enough to prove that in the sensitive axis, the magnetic field and gradient field in low-frequency region will show the PSD with background field different characteristics.

 Converting the above PSD of the sensitive-axis magnetic field and gradient field into ASD and bringing it into Eq.\ref{eq10}, we can get the spectrum of acceleration noise. The TQ program has a shorter arm length and a higher required signal-to-noise ratio compared to LISA, so TQ's inertial sensor noise requirements are higher than LISA's\cite{li2021tianqin}, 

\begin{equation}
S^{1/2}_{a,TQ}=1\times (1+(\frac{f_{c1}}{f})^{2})^{\frac{1}{2}}(1+(\frac{f}{f_{c2}})^{4})^{\frac{1}{2}} \:fm\cdot s^{-2}\cdot Hz^{-1/2}
 \label{eq16}  
\end{equation}
    
\begin{figure}
  \centering
  \includegraphics[width=0.6\textwidth]{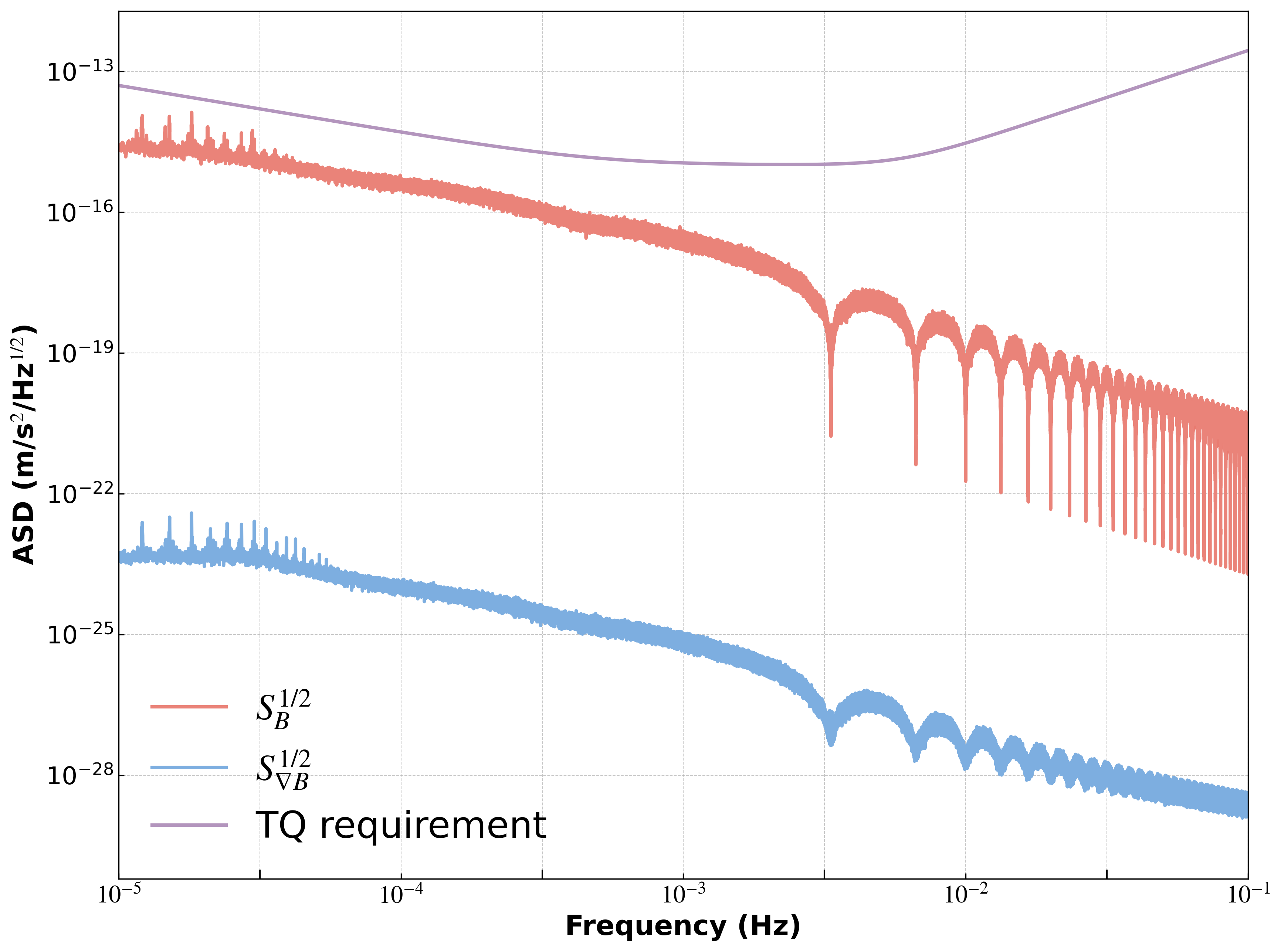}
  \caption{Acceleration noise due to coupling of the magnetic field and magnetic field gradient in the direction of the sensitive axis with the magnetic moment of the test mass in the orbit of the TQ}
  \label{fig:fig4}
\end{figure}

 Where $f_{c1}=0.5mHz$,$f_{c1}=0.5mHz$. Represented by Fig.\ref{fig:fig4}, the red curve represents the acceleration noise spectrum generated by the coupling of the magnetic field to the magnetic moment of the test mass in the direction of the sensitive axis, the blue curve represents the acceleration noise generated by the coupling of the magnetic field gradient to the magnetic moment of the test mass in the direction of the sensitive axis, and the violet curve is the acceleration noise demand of the TQ. Our calculated acceleration noise for the magnetic field and the magnetic field gradient coupled to the remanent magnetic moment of the test mass is lower than the demand curve of the TQ program. We calculate the ratio of the noise to the demand curve, which we define as $\beta$, up to 0.4839, where the effect of the magnetic field gradient coupled acceleration to the test mass is much smaller than the effect of the magnetic field strength.
 
\begin{figure}
  \centering
  \includegraphics[width=0.6\textwidth]{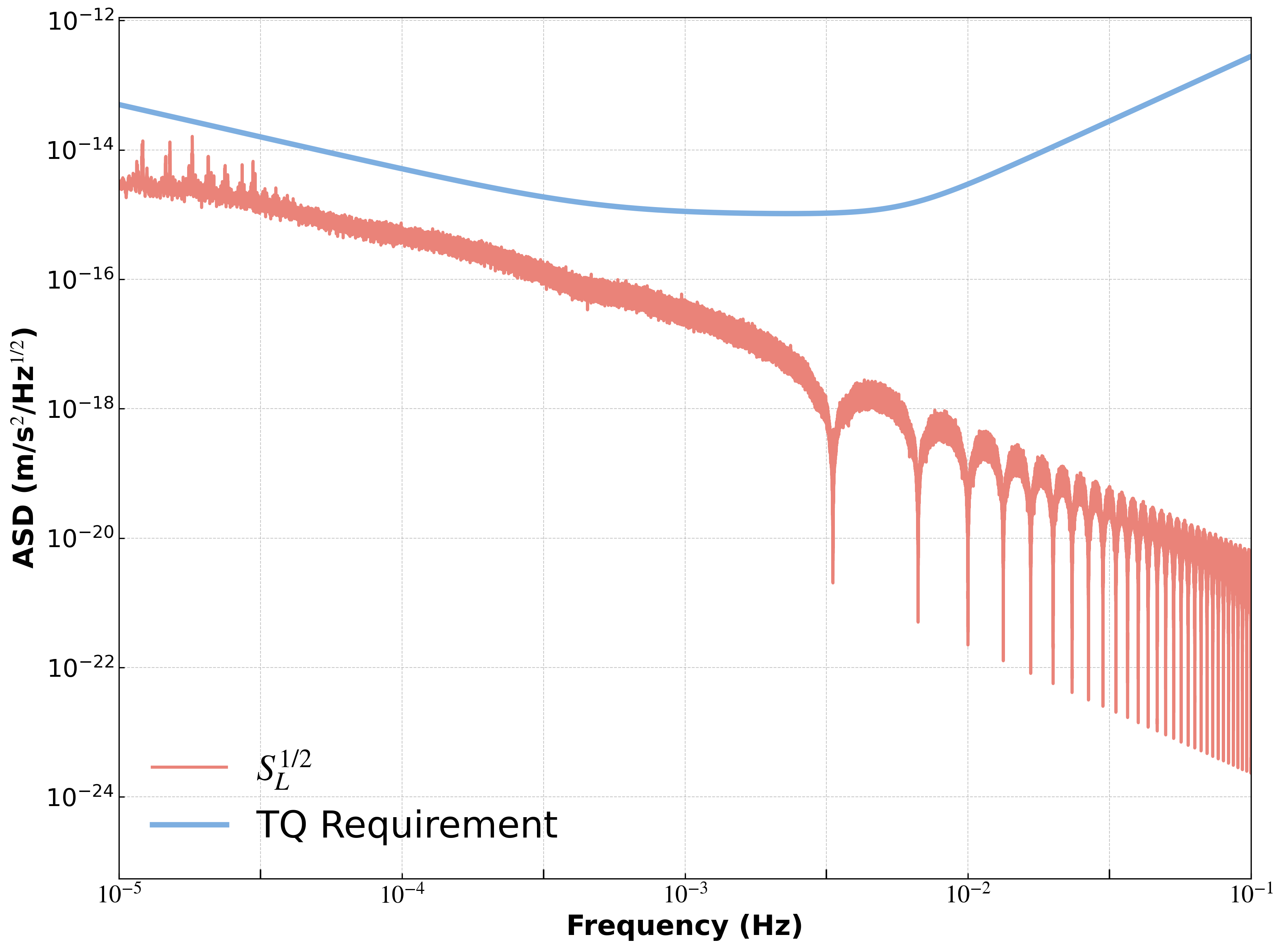}
  \caption{Acceleration noise spectrum resulting from the coupling of the Lorentz force with the space magnetic field}
  \label{fig:fig5}
\end{figure}

\begin{figure}
  \centering
  \includegraphics[width=0.6\textwidth]{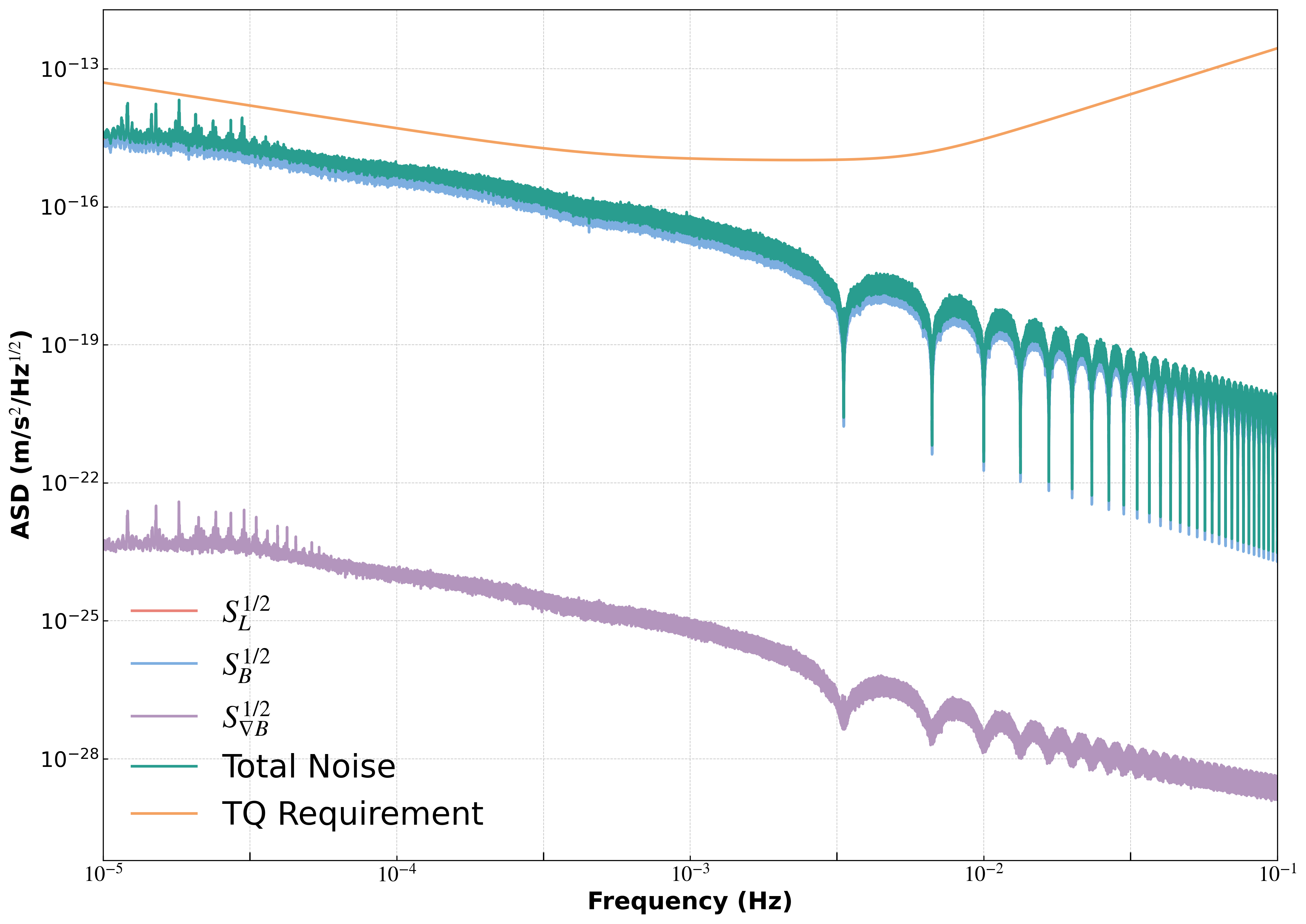}
  \caption{Acceleration noise spectrum generated by all noise terms associated with the space magnetic field}
  \label{fig:fig6}
\end{figure}

 We then bring the PSD of the magnetic and gradient fields in the sensitive axis into Eq.\ref{eq13}. We get the acceleration noise generated by the coupling of the Lorentz force with the magnetic field in space in the sensitive axis, as shown in Fig.\ref{fig:fig5}. We can see that the acceleration noise generated by our calculated Lorentz force can reach $1\times 10^{-15}\:m\cdot s^{-2}\cdot Hz^{-1/2}$. Comparing the Lorentz force-influenced noise with the magnetic field force-influenced noise, as shown in Fig.\ref{fig:fig6}, we can see that the acceleration noise generated by the Lorentz force (the red solid line in Fig.\ref{fig:fig6}, which is covered by the green solid line and the blue realization, as can be seen from the low frequencies, which can be seen from Fig.\ref{fig:fig5}) is about as significant as the acceleration noise generated by the magnetic field (the blue solid line in Fig.\ref{fig:fig6}) and is much larger than the acceleration noise generated by the gradient of the magnetic field acceleration noise (purple solid line in Fig.\ref{fig:fig6}). We synthesize the three noises to obtain the total acceleration noise (green solid line in Fig.\ref{fig:fig6}). We can see that the total acceleration noise is smaller than the acceleration noise profile demanded by TQ,$\beta$ up to 0.7591.

\begin{figure}
  \centering
  \includegraphics[width=0.6\textwidth]{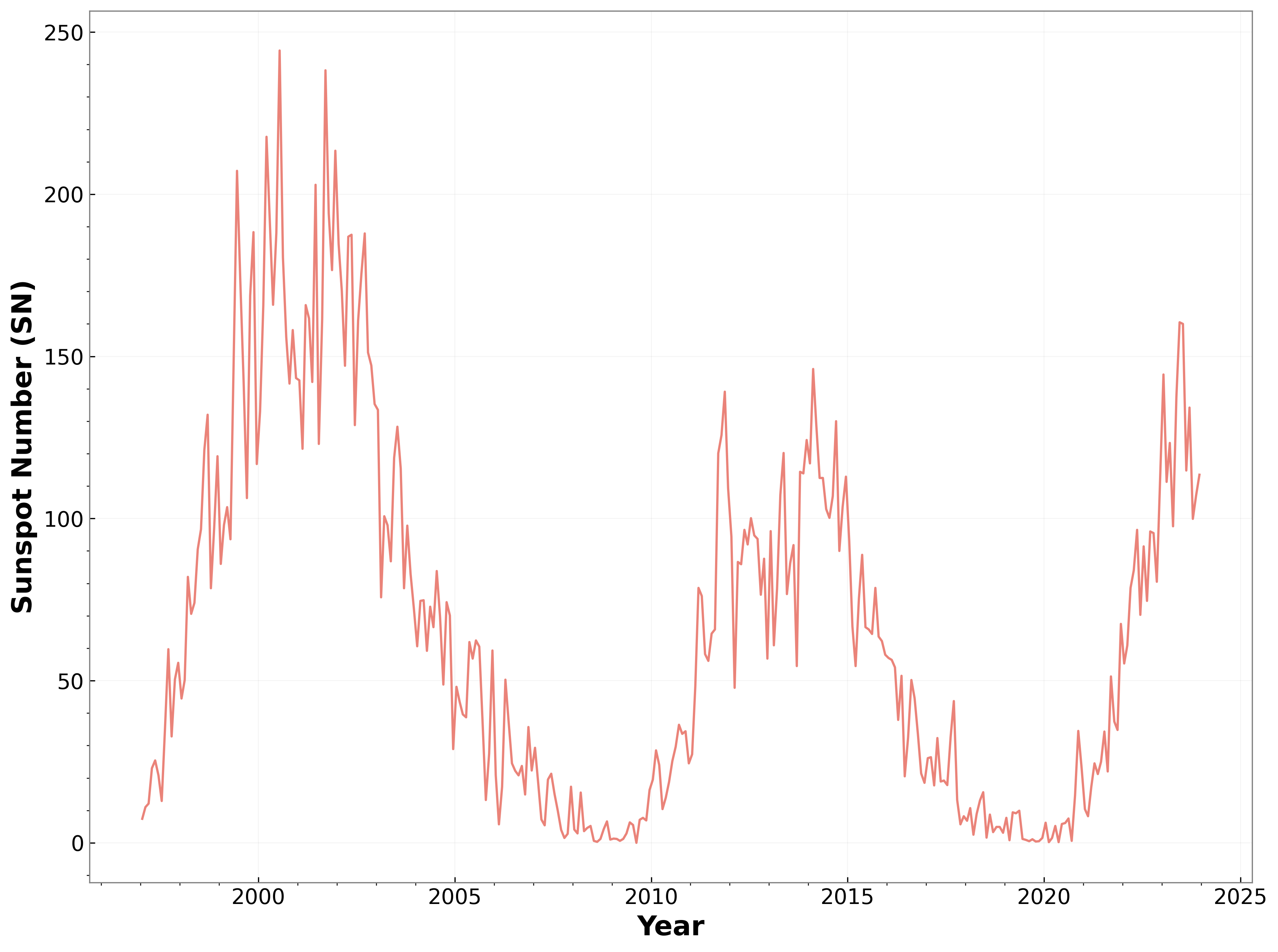}
  \caption{Variation curve of sunspot number with time}
  \label{fig:fig7}
\end{figure}

 In order to investigate the effect of solar activity on the test mass, we extracted the solar activity data during the maximum and minimum periods for calculation. From Fig.\ref{fig:fig7}, we can see that the sunspot book from June 1998 to June 2003 is more than 50, and the sunspot number from January 2006 to December 2010 is less than 30. Therefore, we select the magnetic field data from June 1998 to June 2003 as the data for the considerable period, and the magnetic field data from January 2006 to December 2010 as the data for the minimal period.
 
\begin{figure}
  \centering
  \includegraphics[width=0.6\textwidth]{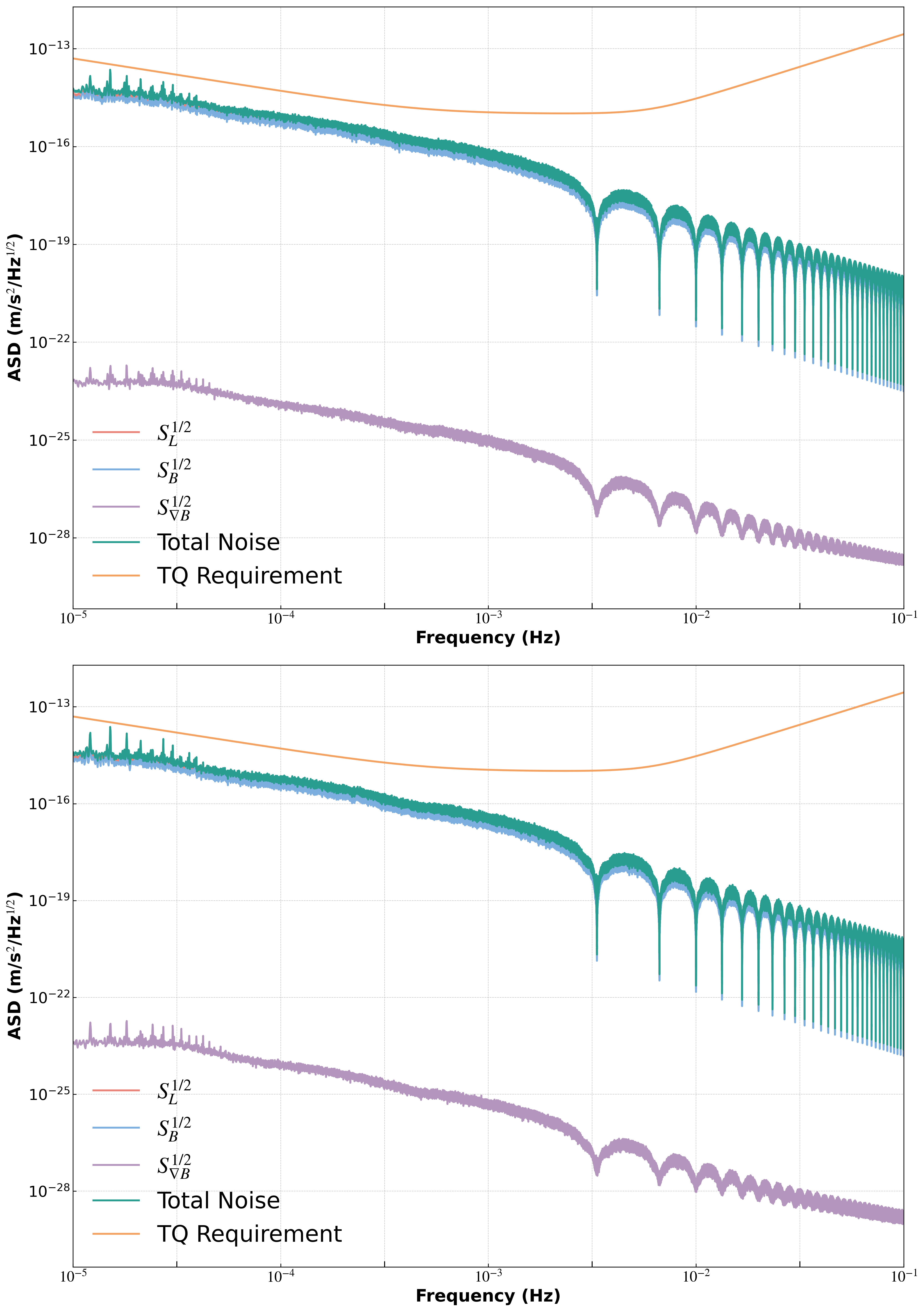}
  \caption{Acceleration noise spectra generated by the space magnetic field corresponding to the period of extreme solar activity and the period of minimal solar activity (the above figure shows the acceleration noise spectrum from June 1998 to June 2003 during the period of extreme solar activity, and the following figure shows the acceleration noise spectrum from January 2006 to December 2010 during the period of minimal solar activity; the green solid line and the blue solid line cover the red solid line in the two figures).}
  \label{fig:fig8}
\end{figure}

 We take the data of the solar activity in the extreme period and the minimal period into Eqs.\ref{eq14} and \ref{eq15} to calculate the magnetic field gradient tensor, and then, according to the relevant data of the celestial orbits, we calculate the magnetic field and the magnetic field gradient of the sensitive axis and convert them into PSD, and take them into Eqs.\ref{eq10} and \ref{eq13}, so that we can calculate the acceleration noise spectra of the magnetic field of the space during the extreme period of the solar activity in the minimal period, as shown in Fig.\ref{fig:fig8}. As shown in Fig.\ref{fig:fig8}, the above figure is the acceleration noise spectrum in the extreme period of solar activity. The following figure is the acceleration noise spectrum in the minimal period of solar activity. After calculation, the maximum value of $\beta$ in the extreme period is 0.6812. The maximum value of $\beta$ in the minimal period is 0.7190. In addition, we can see that in extreme and minimal periods, the acceleration noise is still lower than that of the acceleration noise curve required by the celestial organ. In addition, we can see that the acceleration noise value of the considerable period is higher than that of the minimal period, corresponding to the effect of solar activity. When the solar activity is intense, the solar wind dynamic pressure is enhanced, and the direction of the Earth's magnetosphere facing the Sun is compressed, which will lead to the enhancement of the local magnetic field, and therefore lead to a corresponding increase in the acceleration noise caused by the magnetic field in the considerable period.

\subsection{Discussion}

 Because the TQ orbit is in a 100,000-km geocentric orbit, and the orbit is still located in the Earth's magnetosphere most of the time due to the changes of the Earth's magnetosphere, we have chosen the TA16 model as the model of our space magnetic field. One point should be noted: in the TA16 modelling process, the grid of RBF is only up to 14.5Re, and the farther region needs to be extrapolated. Still, our calculated orbit is at about 15.5Re, which is only about 1Re away from the farthest grid, compared to the distance between the outermost node and the next outermost node, which is about 3Re. This treatment is feasible.
 
 When we calculated the magnetic field, because the minimum time resolution of the current parameters of the TA16 model is 5 minutes, we interpolated the magnetic field to take the frequency to $1\times 10^{-5}$~$1\times 10^{-1}$Hz. Such interpolation will make the PSD of the magnetic field lose the high-frequency features, so the PSD plot presents an oscillatory state in the high-frequency part. Also, when we calculated the magnetic moment, we ignored the shape characteristics of the test mass. We thought the magnetic moment is uniform, not in line with the actual situation. Still, we think that considering the influence of the shape will only affect the change of the value and will not cause the change of the magnitude, so we still think that the magnetic moment is uniform.

\section{Summary}

In this paper, we use the RBF-based TA16 magnetic field model of the space magnetic field to calculate the magnetic field and magnetic field gradient on the celestial TQ orbit, combined with the residual and induced magnetic moments of the test mass, the acceleration noise due to the magnetization is calculated, after that, we also calculate the acceleration noise about the Lorentz force and compare it with the magnetization acceleration noise, the calculations show that the gradient of the magnetic field produces the acceleration noise is much lower than the acceleration noise due to magnetic field and Lorentz force. The total acceleration noise was also calculated, and it can be seen from the results that it fully meets the programed needs of the TQ program. In order to investigate the effect of extreme solar activity on the acceleration noise, we took the acceleration noise spectra from June 1998 to June 2003 and from January 2006 to December 2010. We found that the acceleration noise spectra of the extreme solar activity are slightly higher than those of the non-extreme solar activity.

\section*{Acknowledgments}
This work was supported by the National Key R\&D Program of China (Task No. 2022YFC2204101)

\bibliographystyle{unsrt}  
\bibliography{paper}

\begin{thebibliography}{10}

\bibitem{luo2019overall}
Ziren Luo, Min Zhang, and Gang Jin.
\newblock Overall discussion on the key problems of a space-borne laser interferometer gravitational wave antenna.
\newblock {\em Chinese Science Bulletin}, 64(24):2468--2474, 2019.

\bibitem{Aasi_2015}
The LIGO~Scientific Collaboration, J~Aasi, B~P Abbott, R~Abbott, et~al.
\newblock Advanced ligo.
\newblock {\em Classical and Quantum Gravity}, 32(7):074001, mar 2015.

\bibitem{PhysRevX.9.031040}
B.~P. Abbott, R.~Abbott, T.~D. Abbott, S.~Abraham, et~al.
\newblock Gwtc-1: A gravitational-wave transient catalog of compact binary mergers observed by ligo and virgo during the first and second observing runs.
\newblock {\em Phys. Rev. X}, 9:031040, Sep 2019.

\bibitem{10.1093/ptep/ptaa125}
T~Akutsu, M~Ando, K~Arai, et~al.
\newblock Overview of kagra: Detector design and construction history.
\newblock {\em Progress of Theoretical and Experimental Physics}, 2021(5):05A101, 08 2020.

\bibitem{auclair2023cosmology}
P.~Auclair, D.~Bacon, T.~Baker, et~al.
\newblock Cosmology with the laser interferometer space antenna.
\newblock {\em Living Reviews in Relativity}, 26(5), 2023.
\newblock Received: 19 April 2022; Accepted: 13 June 2023; Published: 28 August 2023.

\bibitem{torres2024detection}
Alejandro Torres-Orjuela, Shun-Jia Huang, Zheng-Cheng Liang, Shuai Liu, Hai-Tian Wang, Chang-Qing Ye, Yi-Ming Hu, and Jianwei Mei.
\newblock Detection of astrophysical gravitational wave sources by tianqin and lisa.
\newblock {\em SCIENCE CHINA Physics, Mechanics \& Astronomy}, 67(5):259511, 2024.

\bibitem{doi:10.1142/S0217751X2050075X}
Wen-Hong Ruan, Zong-Kuan Guo, Rong-Gen Cai, and Yuan-Zhong Zhang.
\newblock Taiji program: Gravitational-wave sources.
\newblock {\em International Journal of Modern Physics A}, 35(17):2050075, 2020.

\bibitem{10.1093/ptep/ptab019}
Seiji Kawamura, Masaki Ando, Naoki Seto, et~al.
\newblock Current status of space gravitational wave antenna decigo and b-decigo.
\newblock {\em Progress of Theoretical and Experimental Physics}, 2021(5):05A105, 02 2021.

\bibitem{s24237685}
Shaoxin Wang, Dongxu Liu, Xuan Zhan, Peng Dong, Jia Shen, Juan Wang, Ruihong Gao, Weichuan Guo, Peng Xu, Keqi Qi, and Ziren Luo.
\newblock Core payload of the space gravitational wave observatory: Inertial sensor and its critical technologies.
\newblock {\em Sensors}, 24(23), 2024.

\bibitem{https://doi.org/10.1029/2020JA028579}
Ling-Feng Lu, Wei Su, Xuefeng Zhang, Zhao-Guo He, Hui-Zong Duan, Yuan-Ze Jiang, and Hsien-Chi Yeh.
\newblock Effects of the space plasma density oscillation on the interspacecraft laser ranging for tianqin gravitational wave observatory.
\newblock {\em Journal of Geophysical Research: Space Physics}, 126(2):e2020JA028579, 2021.
\newblock e2020JA028579 2020JA028579.

\bibitem{https://doi.org/10.1029/93JA01150}
Mauricio Peredo, David~P. Stern, and Nikolai~A. Tsyganenko.
\newblock Are existing magnetospheric models excessively stretched?
\newblock {\em Journal of Geophysical Research: Space Physics}, 98(A9):15343--15354, 1993.

\bibitem{TSYGANENKO19895}
N.A. Tsyganenko.
\newblock A magnetospheric magnetic field model with a warped tail current sheet.
\newblock {\em Planetary and Space Science}, 37(1):5--20, 1989.

\bibitem{gong2021concepts}
Y.~Gong, J.~Luo, and B.~Wang.
\newblock Concepts and status of chinese space gravitational wave detection projects.
\newblock {\em Nature Astronomy}, 5:881--889, 2021.
\newblock Received: 05 April 2021; Accepted: 28 July 2021; Published: 15 September 2021.

\bibitem{Su_2021}
Wei Su, Yan Wang, Chen Zhou, Lingfeng Lu, Ze-Bing Zhou, T.~M. Li, Tong Shi, Xin-Chun Hu, Ming-Yue Zhou, Ming Wang, Hsien-Chi Yeh, Han Wang, and P.~F. Chen.
\newblock Analyses of laser propagation noises for tianqin gravitational wave observatory based on the global magnetosphere mhd simulations.
\newblock {\em The Astrophysical Journal}, 914(2):139, jun 2021.

\bibitem{Luo_2016}
Jun Luo, Li-Sheng Chen, Hui-Zong Duan, Yun-Gui Gong, Shoucun Hu, Jianghui Ji, Qi~Liu, Jianwei Mei, Vadim Milyukov, Mikhail Sazhin, Cheng-Gang Shao, Viktor~T Toth, Hai-Bo Tu, Yamin Wang, Yan Wang, Hsien-Chi Yeh, Ming-Sheng Zhan, Yonghe Zhang, Vladimir Zharov, and Ze-Bing Zhou.
\newblock Tianqin: a space-borne gravitational wave detector.
\newblock {\em Classical and Quantum Gravity}, 33(3):035010, jan 2016.

\bibitem{10.1093/ptep/ptaa114}
Jianwei Mei, Yan-Zheng Bai, Jiahui Bao, et~al.
\newblock The tianqin project: Current progress on science and technology.
\newblock {\em Progress of Theoretical and Experimental Physics}, 2021(5):05A107, 08 2020.

\bibitem{Su_2020}
Wei Su, Yan Wang, Ze-Bing Zhou, Yan-Zheng Bai, Yang Guo, Chen Zhou, Tom Lee, Ming Wang, Ming-Yue Zhou, Tong Shi, Hang Yin, and Bu-Tian Zhang.
\newblock Analyses of residual accelerations for tianqin based on the global mhd simulation.
\newblock {\em Classical and Quantum Gravity}, 37(18):185017, aug 2020.

\bibitem{PhysRevD.108.103030}
Wei Su, Ze-Bing Zhou, Yan Wang, Chen Zhou, P.~F. Chen, Wei Hong, J.~H. Peng, Yun Yang, and Y.~W. Ni.
\newblock Evaluating residual acceleration noise for the tianqin gravitational waves observatory with an empirical magnetic field model.
\newblock {\em Phys. Rev. D}, 108:103030, Nov 2023.

\bibitem{https://doi.org/10.1002/2016JA023217}
N.~A. Tsyganenko and V.~A. Andreeva.
\newblock An empirical rbf model of the magnetosphere parameterized by interplanetary and ground-based drivers.
\newblock {\em Journal of Geophysical Research: Space Physics}, 121(11):10,786--10,802, 2016.

\bibitem{milyukov2020tianqin}
V.~K. Milyukov.
\newblock Tianqin space-based gravitational wave detector: Key technologies and current state of implementation.
\newblock {\em Astronomical Reports}, 64:1067--1077, 2020.
\newblock Received: 15 June 2020; Revised: 29 June 2020; Accepted: 30 June 2020; Published: 30 December 2020.

\bibitem{https://doi.org/10.1029/RG014i002p00199}
David~P. Stern.
\newblock Representation of magnetic fields in space.
\newblock {\em Reviews of Geophysics}, 14(2):199--214, 1976.

\bibitem{https://doi.org/10.1029/2012JA018056}
Ingrid Cnossen and Arthur~D. Richmond.
\newblock How changes in the tilt angle of the geomagnetic dipole affect the coupled magnetosphere-ionosphere-thermosphere system.
\newblock {\em Journal of Geophysical Research: Space Physics}, 117(A10), 2012.

\bibitem{TSYGANENKO19871347}
N.A. Tsyganenko.
\newblock Global quantitative models of the geomagnetic field in the cislunar magnetosphere for different disturbance levels.
\newblock {\em Planetary and Space Science}, 35(11):1347--1358, 1987.

\bibitem{https://doi.org/10.1029/2001JA000219}
N.~A. Tsyganenko.
\newblock A model of the near magnetosphere with a dawn-dusk asymmetry 1. mathematical structure.
\newblock {\em Journal of Geophysical Research: Space Physics}, 107(A8):SMP 12--1--SMP 12--15, 2002.

\bibitem{https://doi.org/10.1029/2001JA000220}
N.~A. Tsyganenko.
\newblock A model of the near magnetosphere with a dawn-dusk asymmetry 2. parameterization and fitting to observations.
\newblock {\em Journal of Geophysical Research: Space Physics}, 107(A8):SMP 10--1--SMP 10--17, 2002.

\bibitem{https://doi.org/10.1002/2015JA021641}
N.~A. Tsyganenko and V.~A. Andreeva.
\newblock A forecasting model of the magnetosphere driven by an optimal solar wind coupling function.
\newblock {\em Journal of Geophysical Research: Space Physics}, 120(10):8401--8425, 2015.

\bibitem{BORODKOVA20081220}
N.L. Borodkova, J.B. Liu, Z.H. Huang, and G.N. Zastenker.
\newblock Geosynchronous magnetic field response to the large and fast solar wind dynamic pressure change.
\newblock {\em Advances in Space Research}, 41(8):1220--1225, 2008.

\bibitem{PhysRevD.97.122002}
M.~Armano, H.~Audley, J.~Baird, et~al.
\newblock Calibrating the system dynamics of lisa pathfinder.
\newblock {\em Phys. Rev. D}, 97:122002, Jun 2018.

\bibitem{Sumner_2020}
Timothy~J Sumner, Guido Mueller, John~W Conklin, Peter~J Wass, and Daniel Hollington.
\newblock Charge induced acceleration noise in the lisa gravitational reference sensor.
\newblock {\em Classical and Quantum Gravity}, 37(4):045010, jan 2020.

\bibitem{li2021tianqin}
Hongyin Li, Yanchong Liu, Chengrui Wang, Y.~Bai, L.~Liu, S.~Wu, S.~Qu, D.~Tan, H.~Yin, Z.~Li, S.~Yang, and Z.~Zhou.
\newblock Preliminary design considerations and progress of the tianqin inertial sensor.
\newblock {\em Acta Scientiarum Naturalium Universitatis Sunyatseni}, 60(Suppl. 1), 2021.

\end{thebibliography}

\end{document}